\documentstyle[11pt,newpasp,twoside,epsf,natbib]{article}

\begin{document}
\title{Investigating the long-term evolution of galaxies: \\
  Noise, cuspy halos and bars} 
\author{Martin D. Weinberg} 
\affil{University of Massachusetts, Amherst, Massachusetts, USA}

\begin{abstract}
  I review the arguments for the importance of halo structure in
  driving galaxy evolution and coupling a galaxy to its environment.
  We begin with a general discussion of the key dynamics and examples
  of structure dominated by modes.  We find that simulations with
  large numbers of particles ($N\ga10^6$) are required to resolve the
  dynamics.  Finally, I will describe some new results which
  demonstrates that a disk bar can produce cores in a cuspy CDM
  dark-matter profile within a gigayear.  An inner Lindblad-like
  resonance couples the rotating bar to halo orbits at all radii
  through the cusp, rapidly flattening it.  This resonance disappears
  for profiles with cores and is responsible for a qualitative
  difference in bar-driven halo evolution with and without a cusp.
  Although the bar gives up the angular momentum in its pattern to
  make the core, the formation epoch is rich in accretion events to
  recreate or trigger a classic stellar bar. The evolution of the
  cuspy inner halo by the first-generation bar paves the way for a
  long-lived subsequent bar with low torque and a stable pattern
  speed.
\end{abstract}

\section{Introduction}

Enormous effort in simulating the evolution of large scale structure
over the last five years has led to a convincing scenario for galaxy
formation and the early phases of evolution.  Although these
simulations are very far from a complete description of ISM and star
formation, many of the basic features our in concert with
observations.  The final frontier is an understanding of evolution
from formation to present.  This is a very difficult problem for a
number of reasons.  First, the time scales are relatively long and
span many characteristic orbital times.  Second, there are multiple
length scales involved: disk scale heights, scale lengths, dark-matter
scale lengths.  The evolution of galaxies almost certainly depends on
the interaction between these scales.

My goal for this talk is to give you my views on the essential
features that must be borne in mind when approaching this problem
based on my own work and the work of others.  I will emphasize stellar
dynamics and implications for simulations.  The talk will have two
parts parts.  I will begin with a review of the basic dynamics of
galaxy structure (\S\ref{sec:dynamics}). It is natural to think in
terms of solving Newton's equations for $N$ particles.  As long as
this is done accurately, one only needs to interpret the simulation
output.  Of course, most of you are aware that this is naive and may
easily lead to the wrong results.  In the $N\rightarrow\infty$ limit,
collective modes may dominate the structure of galaxies.  This is
obvious in disks where spirals and bars are apparent but true for
halos as well, as I will describe.  The majority of this talk will
describe two scenarios: noise evolution
(\S\S\ref{sec:noisepower}--\ref{sec:noiseevol}) and bar-driven
evolution\footnote{in collaboration with Neal Katz} (\S\ref{sec:barhalo}).

\section{Key features of stellar systems}
\label{sec:dynamics}

Collisionless dynamics has a modal structure very different from
everyday mechanical systems.  A perfect elastic system (the textbook
case of a drum-head or string) has infinite number of discrete modes.
One may excite one mode at a time in principle.  The situation is
rather different for a stellar system.  In general, stellar systems
have an infinite number of continuous modes.  Any real perturbation
will excite infinitely many at a time.  A wave-packet in a dispersive
medium is a good analogy.  The result is that nearly all perturbations
to a galaxy damp.  A practical example of these dynamics is {\em
  dynamical friction}.  The familiar Chandrasekhar formula computes
the change in momentum by accumulating the gravitational scattering
between the perturbing mass and every other trajectory in the system.
Alternatively, we may consider the response of the system to the
moving body.  The body passing through a stellar medium excites a
wake.  This wake exerts a gravitational tug on the body slowing it
down.  Relative to a particular center, the body's motion may be
Fourier decomposed in to a broad continuous spectrum of frequencies,
which couple to the system's modes at some multiple of these
frequencies.  Small multiples general lead to stronger excitation.
The wake, then, consists of a superposition of continuous modes which
quickly mixes away as the body continues on.

\subsection{Exceptions to damping}

There are some notable exceptions to this strong damping.  For
example, one often sees spiral arms and bars in galactic disks.  Why?
Because the phase space is {\em squeezed} in its degrees of freedom.
It is nearly a two-dimensional system with the vertical motion of
acting independently of the radial and tangential ones.  Similarly,
the velocity dispersion is small compared to the mean tangential
velocity.  One can exploit the reduced dimensionality and disparate
radial, tangential and vertical frequencies to find pattern speed
which only couples weakly to the system's modes.  In other words, the
fewer the degrees of freedom, the fewer opportunities for coupling and
damping.  Bending modes are a good example.  They are undamped for an
infinitesimally thin disk but damped become damped as the thickness
increases \citep{Weinberg:91*1}.  The damping is caused by coupling
between the longitudinal bending and the spectrum modes in with
vertical motion.  The larger the ratio of vertical to radial
dispersion, the more disparate the frequencies and the weaker the
damping.

Now what about the dark-matter halo?  The halo is inhomogeneous in the
radial direction: {\em squeezed} into its own gravitational bottle.
This limits range of possible modal frequencies.  Therefore just as
the disk case, there are pattern speeds which weakly couples to the
halo's continuum modes.  Therefore, a halo can exhibit a weakly damped
response!  Note that this situation is not true for an {\em infinite
  stellar medium}.  The standard result that a {\em hot} stellar
system will not exhibit patterns derives from the infinite stellar
medium and is not correct for a real galaxy.  More on this below.

\subsection{Weakly damped modes in a dark-matter halo}

These weakly damped modes are a natural part of the dynamics for
nearly all halos.  I have found them in every model I have examined so
far, both with and without cusps.  For example, Figures 4 and 5 in
\citet{Weinberg:94} shows the mode shape and total density shape for
an excited mode in a King $W_0=5$ model.  This dipole mode shifts the
center of the halo giving it a lopsided appearance.  The pattern speed
is non-zero but the damping time 1--2 orders of magnitude smaller than
pattern speed!

\begin{figure}
  \plottwo{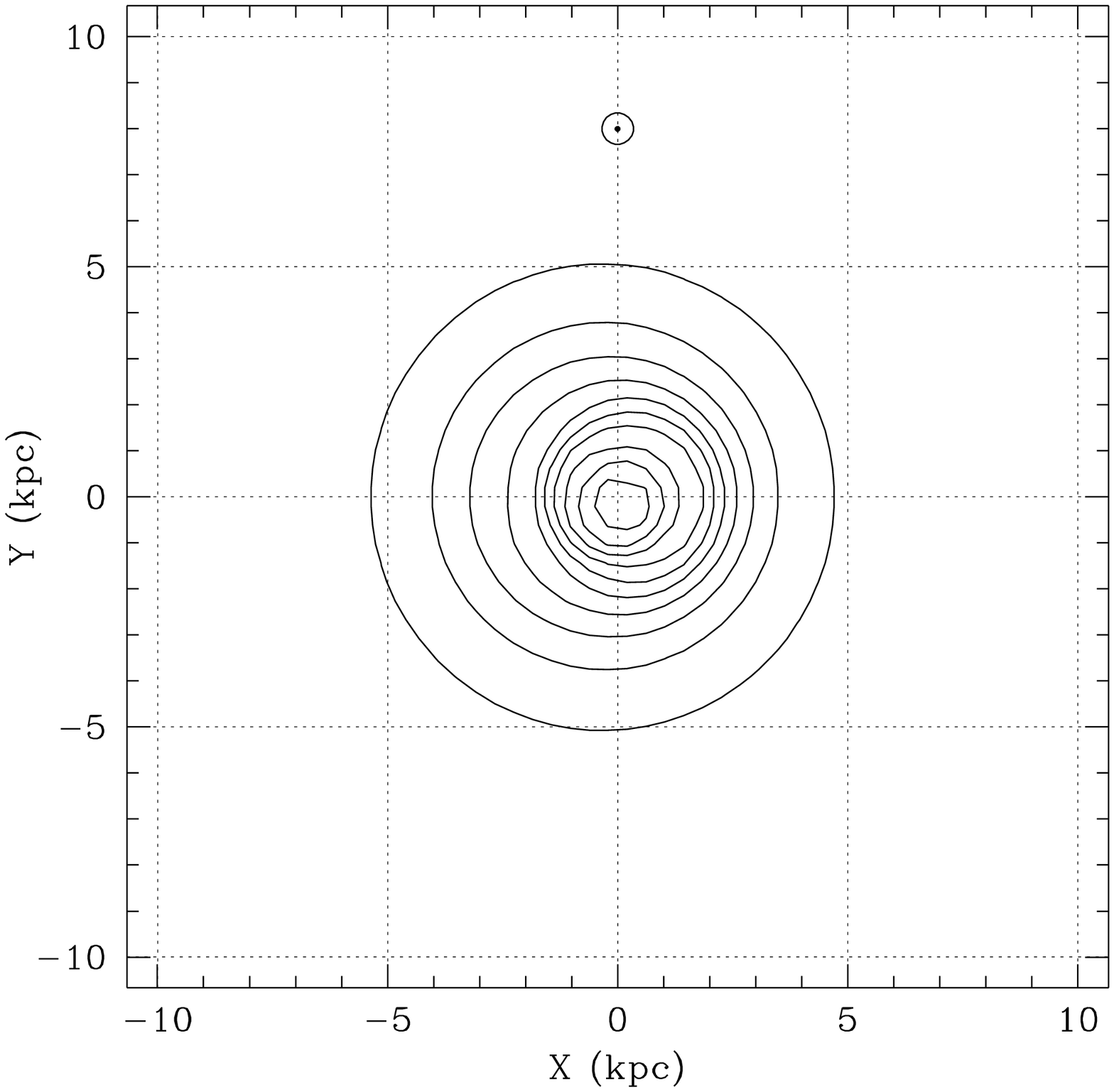}{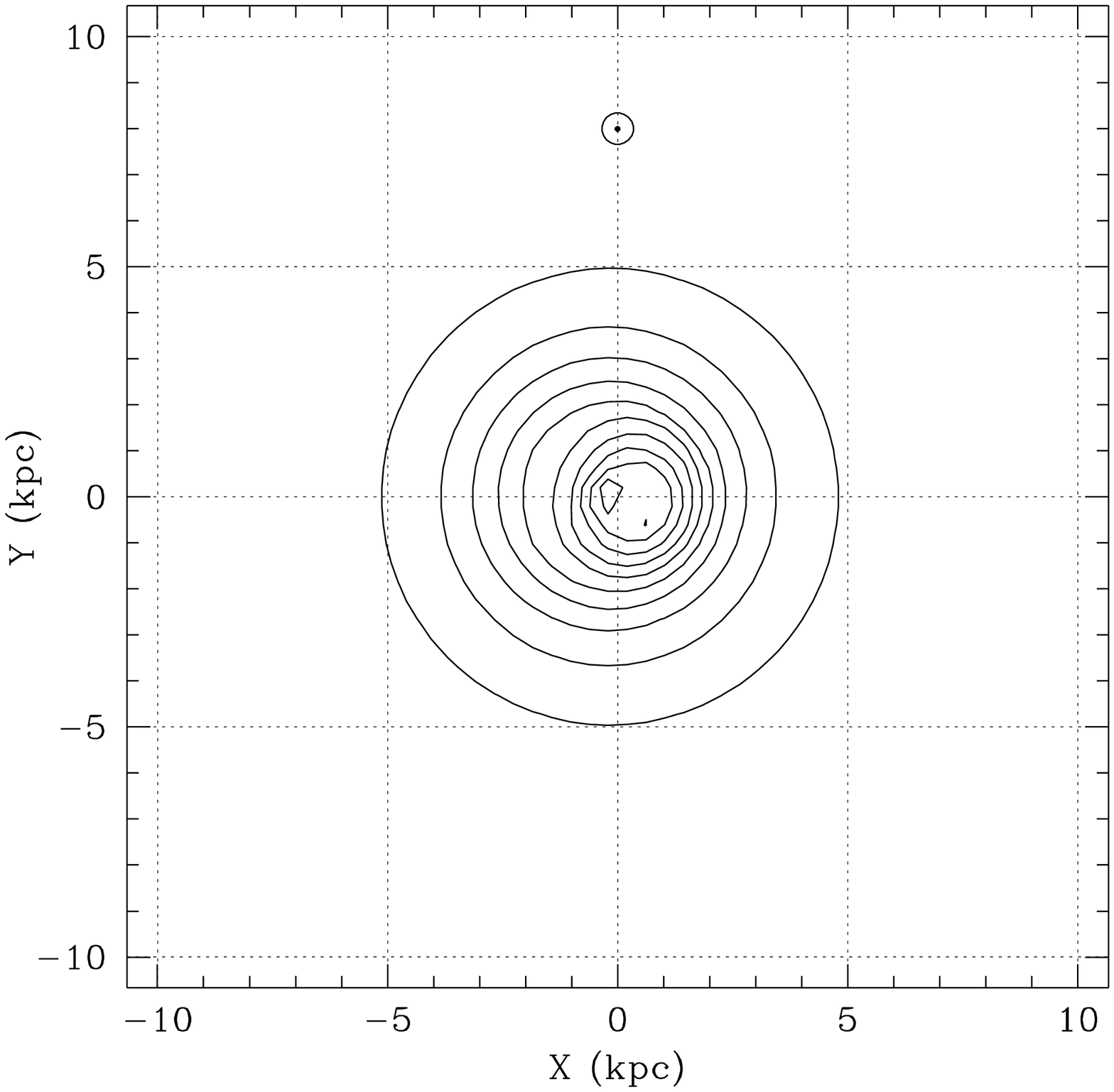}
  \caption{Prograde (left) and retrograde (right) modes for the a
    combined Toomre disk and $W_0=5$ King halo.}
  \label{fig:w05diskhalo}
\end{figure}

It is natural to consider the disk separately in all of this.  Of
course, the disk and the halo are coupled gravitationally and together
the two have a weakly damped modes.  The rotation of the disk breaks
the spherical symmetry and now there is a prograde and retrograde
weakly damped mode.  It won't surprise you to learn that the prograde
mode damps more quickly than the retrograde mode but the shape of the
modes are quite similar (see Fig. \ref{fig:w05diskhalo}).

\subsection{Difficulty resolving modes in simulations}

In order to see these modes in n-body simulations, the noise has to be
small enough that the diffusion time for orbits is large compared to
pass through the resonance (the libration time).  This requires
low-noise simulations.  For this purpose, I use the expansion method
\citep{Clutton-Brock:72,Clutton-Brock:73,Kalnajs:76*3,
  Fridman.Polyachenko:84, Hernquist.Ostriker:92}.  The use of
orthogonal {\em wave functions} effectively filters out noise below
the length scale of interest.  These papers describe bases with
density profiles that may not look like the underlying galaxy.  A good
fit with an arbitrary basis requires more terms in the series.  Each
of these terms introduces another degree of freedom which adds to the
total noise.  \cite{Weinberg:99} shows that one can choose a basis
adaptively to match the equilibrium profile and the series then
converges quickly and will result in low noise.

There are a number of advantages to this Poisson solver.  It is {\em
  embarrassingly parallel} with ${\cal O}(N)$ scaling
\citep[see][]{Hernquist.etal:95}.  The coefficients may be computed on
all processors simultaneously, for example, and therefore has none of
the domain decomposition problems of a tree-code.  Secondly, one can
compare directly with perturbation theory which is one of my main
attractions to this approach.  For these reasons, it is very good
choice for long-term evolution.  The main disadvantage is that the
basis can not adapt to arbitrary morphology so it is not good for
mergers or following the non-linear development of instabilities where
the end result is not a simple known geometry to start.

\section{Disk responding to noisy halo}
\label{sec:noisepower}

Let's look at a simulation of an equilibrium King model halo with an
exponential disk using this method.  The halo concentration is chosen
to match the characteristics of observed rotation curves.  Scaled to
the Milky Way, the halo core radius approximately the solar circle
with a scale length of 3.5 kpc.  The radial velocity dispersion is
chosen large enough to stabilize the disk.  The noise is dominated by
an $m=1$ disturbance and appears as a lopsidedness (as in Fig.
\ref{fig:w05diskhalo}).  The orientation of the distortion varies with
time but with the same shape.  This is due to the excitation of the
damped modes.  The next order contribution to the noise, from $m=2$,
is smaller by an order of magnitude.

\begin{figure}
  \plotfiddle{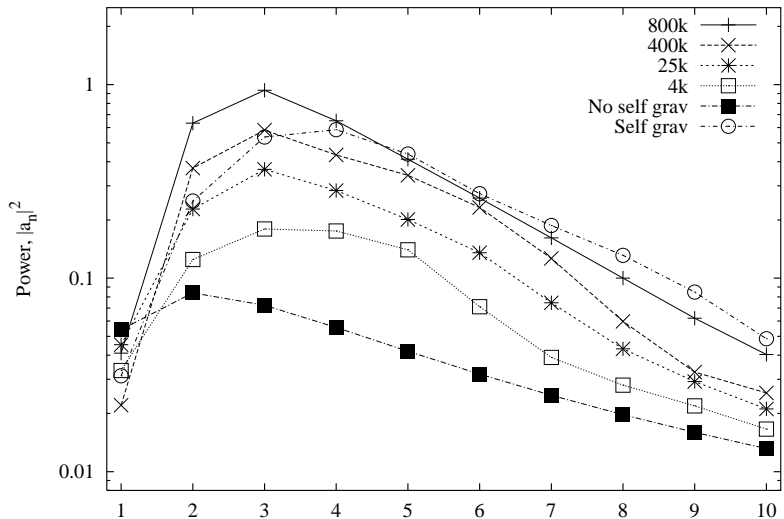}{180pt}{0}{85}{85}{-200pt}{-60pt}
  \caption{Power from particle noise for simulations with varying
    numbers of particle $N$ (labeled).  All are scaled to $N=1$ and
    compared with the result from perturbation theory.}
  \label{fig:powernumber}
\end{figure}

The expected noise amplitude due to excitation of modes can be
computed analytically by perturbation theory and compared with n-body
simulations.  Figure \ref{fig:powernumber} shows the power in each
$l=1$ basis function for a variety of simulations with different $N$
for the halo.  The expected noise without self-gravity (Poisson noise,
in other words) is shown for comparison (filled squares).  For small
$N$, the power in functions with small index $n$ is underestimated
without self-gravity.  The basis function index is equal to the number
nodes; the scale structure is represented primarily by functions with
small index $n$.  At particle number $N\approx10^6$, the simulations
approach the theoretical expectation for small $n$.  I conclude that
one requires $\ga10^6$ particles to recover fine scale structure.  I
have used a low concentration model here; many more particles needed
for higher concentration halos to cover the increased dynamic range.
Also, the expansion method suppresses very small scales and reduce the
total power in noise.  Tree codes and other adaptive schemes
(including direct summation) that can resolve a larger range of scales
pay for this generality by needing a considerably larger $N$ to
achieve the same results.  There is a more general consequence of this
restricted test: the dynamics of the noise-excited modes are
fundamentally the same in any excitation and therefore large particle
numbers will be required to correctly simulate any response (e.g.  a
similar value of $N$ required for convergence for simulations
described in \S\ref{sec:barhalo}).  Conversely, if $N$ is an order of
magnitude below the necessary number, some weakly damped modes may be
completely missing from the response.  Doubling $N$ may not change the
the appearance of the simulation but this will not indicate
convergence.

\section{Evolution of halo profiles by noise}
\label{sec:noiseevol}

The noise described in \S\ref{sec:noisepower} can drive evolution in
the halo.  There are two limiting types of perturbations: point masses
such as massive halo black holes and transient perturbers such as
fly-by encounters, orbital decay, disrupting dwarfs.  Point mass
perturbers only drive halo structure at harmonic order $l>2$ and
self-gravity is less important in this high-$l$ regime\footnote{More
  precisely, any quasi-periodic perturbation falls into this category.
  The only practical case that I can identify is orbital motion.
  However, some arbitrary applied periodic forcing could couple to low
  $l$.}.  The response to transient perturbations is dominated by
$m=1$ ($l=1$) weakly damped modes.  Because the weakly damped modes
dominate the response, its appearance is only weakly dependent on
details of perturbation. This can be demonstrated by direct example;
the same profiles result from evolution by fly-bys and orbitally
decaying point masses.

To compute the evolution, I use methods from stochastic differential
equation.  In short, assuming that individual perturbations are
independent, the system reduces to an equation in Fokker-Planck form.
One solves this Fokker-Planck equation for some time scale larger than
the dynamical time but small compared to overall evolution time scale.
Numerically, the solution procedure is analogous to that for
Fokker-Planck evolution of globular clusters using operator splitting,
etc. \citep[for a review]{Spitzer:87}.

\begin{figure}
  \plottwo{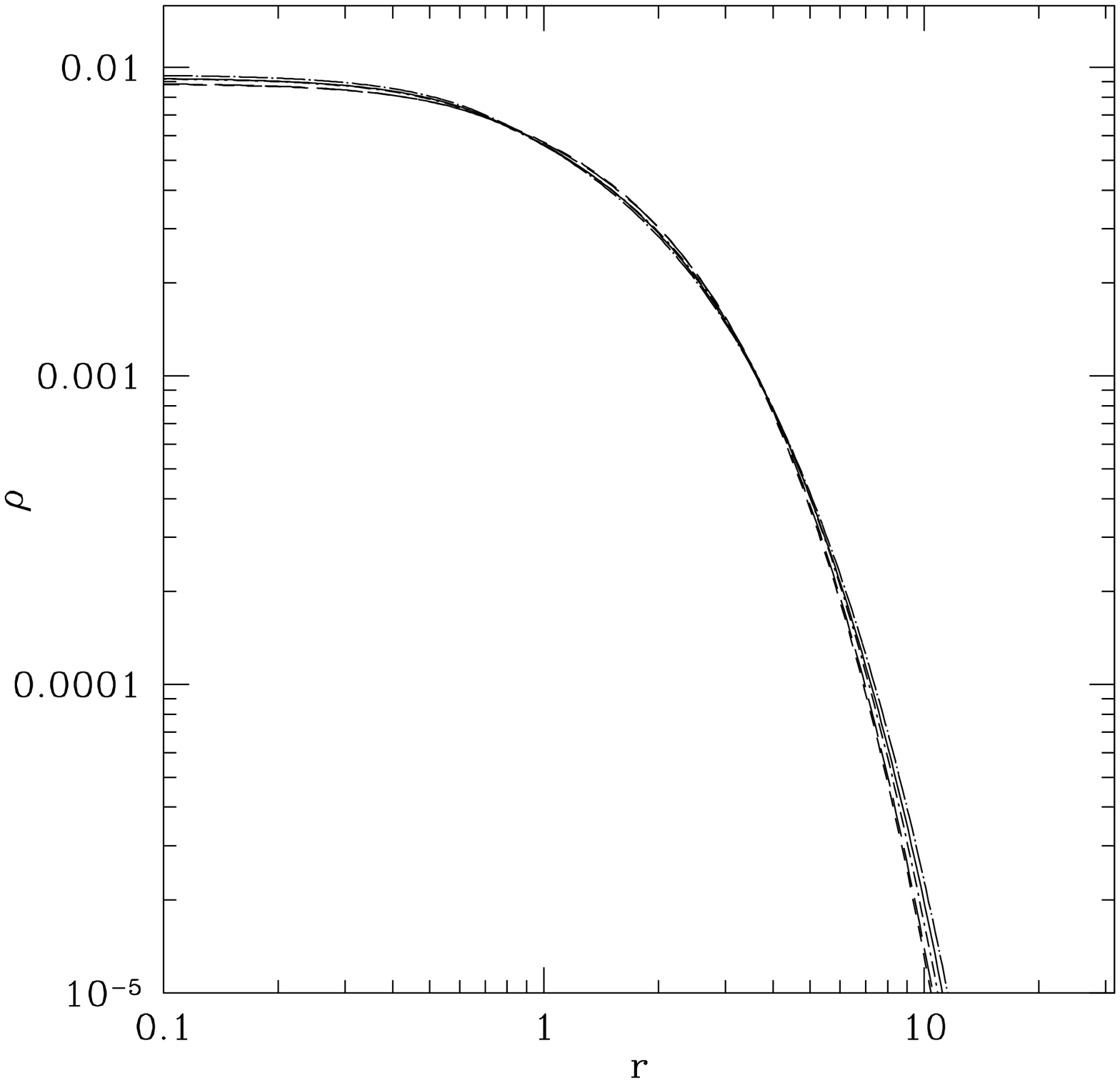}{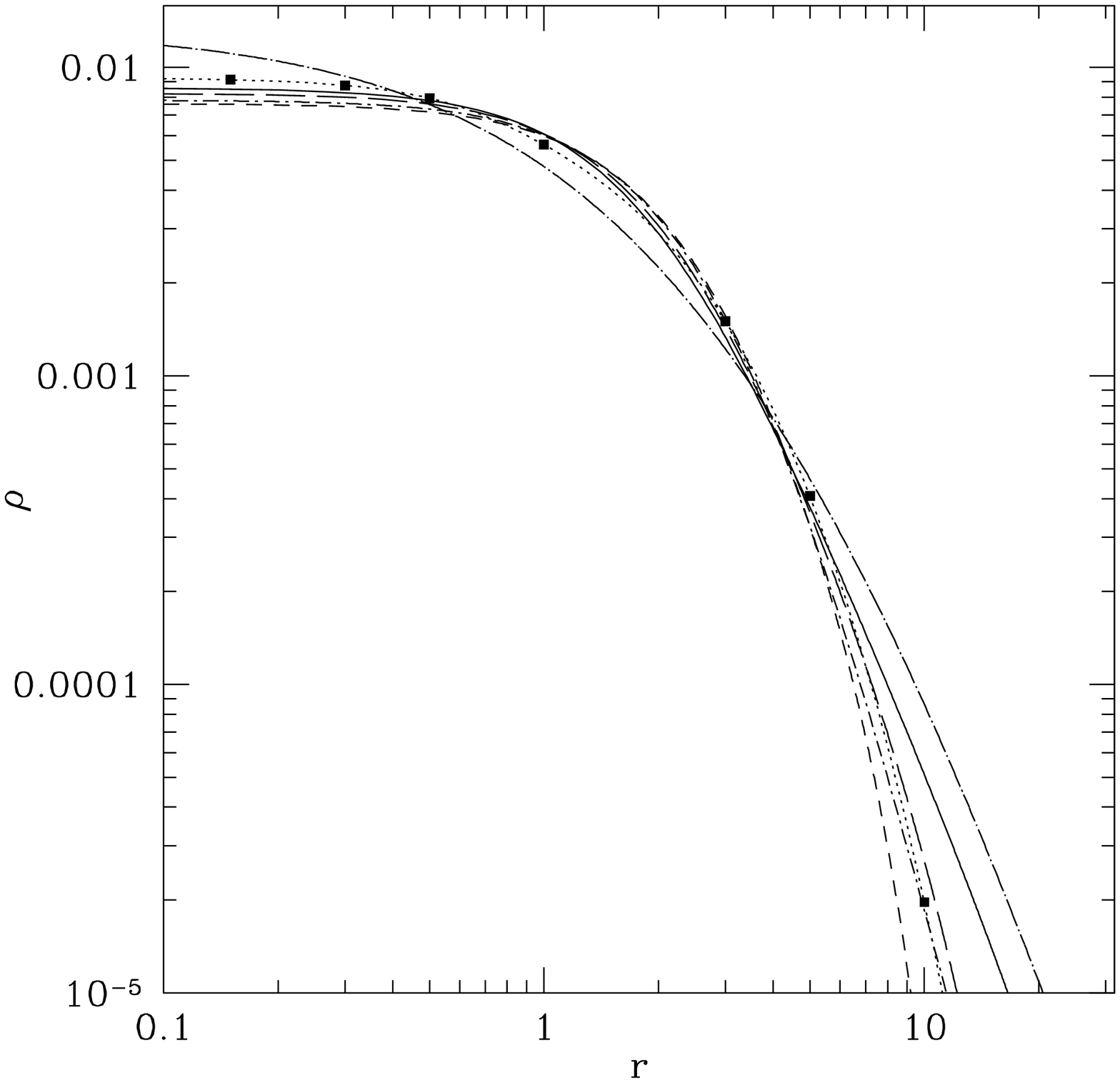}
  \caption{
    Evolved halo profiles for five different initial conditions: King
    models $W_0=3$ (dash), $W_0=5$ (long dash), $W_0=7$ (solid),
    Plummer (dash-dot) and double power law (long dash-dot) with
    $\gamma=1$, $\beta=4$ and $\epsilon=0.1$.  Left: the five profiles
    shown are compared after evolution.  Right: the evolved profile
    are compared with the best fit to the initial profile. (These
    figures originally appeared in {\em Montly Notices of the Royal
      Astronomical Society}.)}
  \label{fig:universal}
\end{figure}

Let's look at some results of evolution due to transient
perturbations.  Figure \ref{fig:universal} shows the halo profile
after evolution due to decaying substructure for a variety of initial
profiles: King models of varying concentration, the Plummer law and
the cold dark matter profile suggested by \citet[][hereafter
NFW]{Navarro.Frenk.White:97}.  As long the $m=1$ modes are excited the
noise looks nearly the same and the profile approaches the same form.
Figure \ref{fig:universal} compares both the evolved profiles and the
fit of the original profiles to the evolved profiles.  The inner part
of the evolved profile can be acceptably fit by King models and
Plummer models (which have cores) but not the NFW profile.  At the
same time, the all evolved profiles are similar, demonstrating the
universality.

\section{Bar-driven evolution}
\label{sec:barhalo}

Cold dark matter (CDM) structure formation simulations predict a
universal cuspy halo profile \citep[NFW]{Navarro.Frenk.White:97}.
This profile, $\rho\propto r^{-\gamma}(1+r/r_s)^{\gamma-3}$ or
$\rho\propto r^{-\gamma}(1+(r/r_s)^{3-\gamma})^{-1}$ with $\gamma=1$,
was first presented by NFW based on a suite of collisionless n-body
simulations with different initial density fluctuation spectra and
cosmological parameters\footnote{Recent work debates the value of
  $\gamma$ \citep[see] {Jing.Suto:00,Moore.etal:98} and but most
  estimates are in the range $1<\gamma<1.5$.}.  Some observational
evidence is consistent with such a cuspy dark matter profile
\citep{vandenBosch.Swaters:01} but not all \citep{deBlok.etal:01}.
Theoretically, disk bars and strong grand-design structure can
strongly interact with the dark matter \citep{Weinberg:85,
  Hernquist.Weinberg:92} by transferring angular momentum to the
spheroid-halo component.  This has been recently elaborated in a pair
of contributions by \citet{Debattista.Sellwood:98,
  Debattista.Sellwood:00}.  They constructed a strong bar from an
Q-unstable disk and following its evolution in a live halo.  They
found that the bars rapidly transferred angular momentum to massive
(non-rotating) halos as predicted by \citet{Weinberg:85}.  However, a
few direct and wider variety of indirect observational inferences
suggest rapidly rotating bars.  Debattista \& Sellwood conclude that
dark-matter halos must have low density (or large cores) to be
consistent with observations.  This failure has potentially serious
implications for the otherwise successful large-scale-structure
paradigm.

After the first of these papers, \citet{Tremaine.Ostriker:99}
hypothesized that the interaction between tidal streamers and disk may
flatten and spin up the inner halo-spheroid.  The drag on bar reduced
from flattened, rotating inner halo is then reduced.  Another
possibility is that there is something wrong with the dark matter
hypothesis itself, perhaps requiring new physics
\citep{Spergel.Steinhardt:00}.

Although this controversy has been illustrated by interesting and
novel possibilities, dynamical evolution offers a natural solution in
the context of CDM structure formation.  In particular, simulations
commonly produce bars in the gas-rich environment of a forming disk.
The same torque which drives the slow down described by
\citet{Weinberg:85} and \citet{Debattista.Sellwood:98} a provides
sufficient torque to remove the inner cusp of an NFW profile soon
after formation.  At first look, this solution appears to run afoul
with the observational evidence for rapid pattern speeds that began
the controversy.  However, this early gas rich bar is not the direct
progenitor of the stellar bars seen in the present epoch.  Moreover,
the removal of the cusp paves the way for the later formation of a
{\em classic} stellar bar with a stable or slowly decreasing pattern
speed.

\subsection{Physical description of the dynamics}
\label{sec:bardynamics}

\begin{figure}
  \plottwo{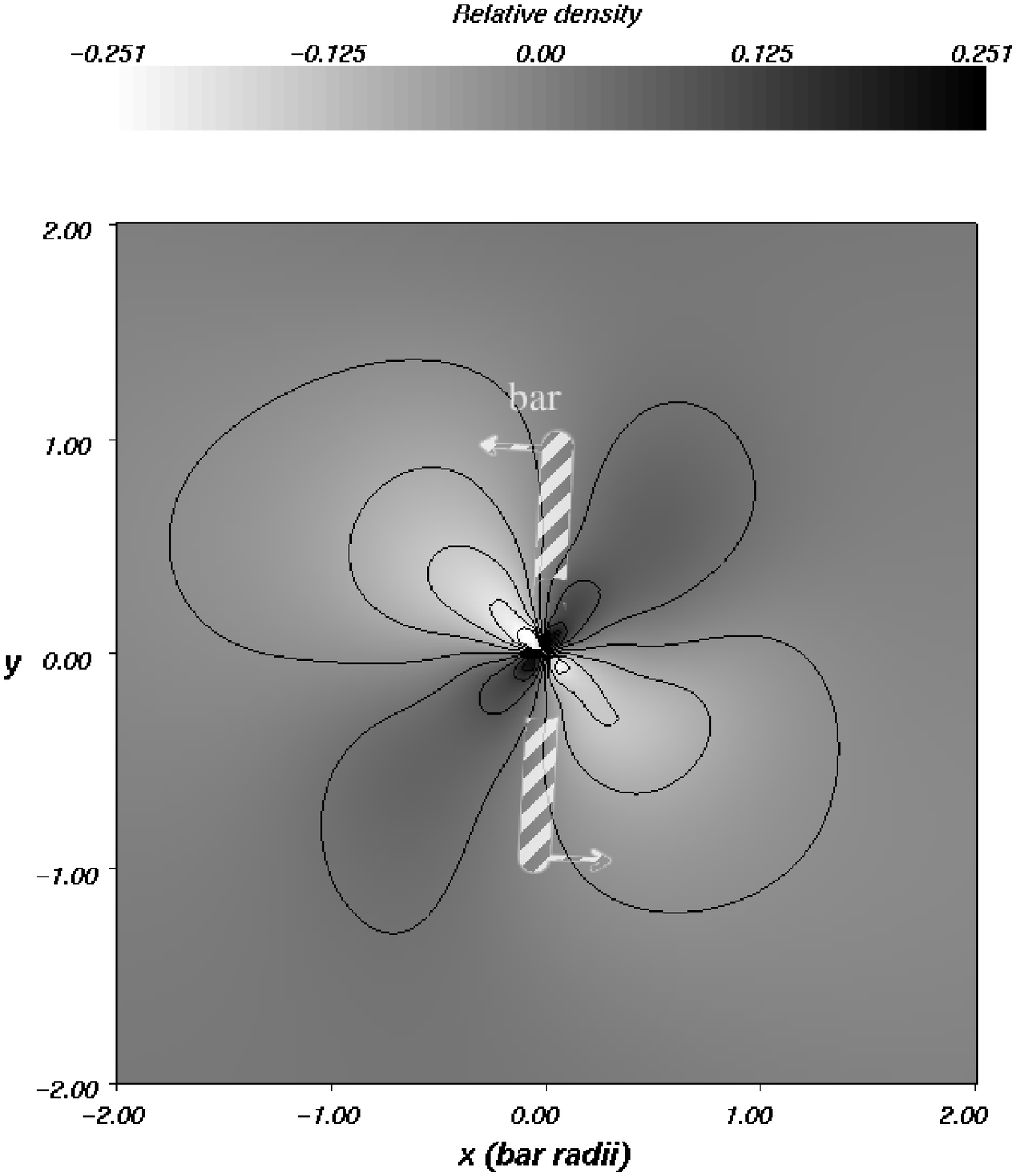}{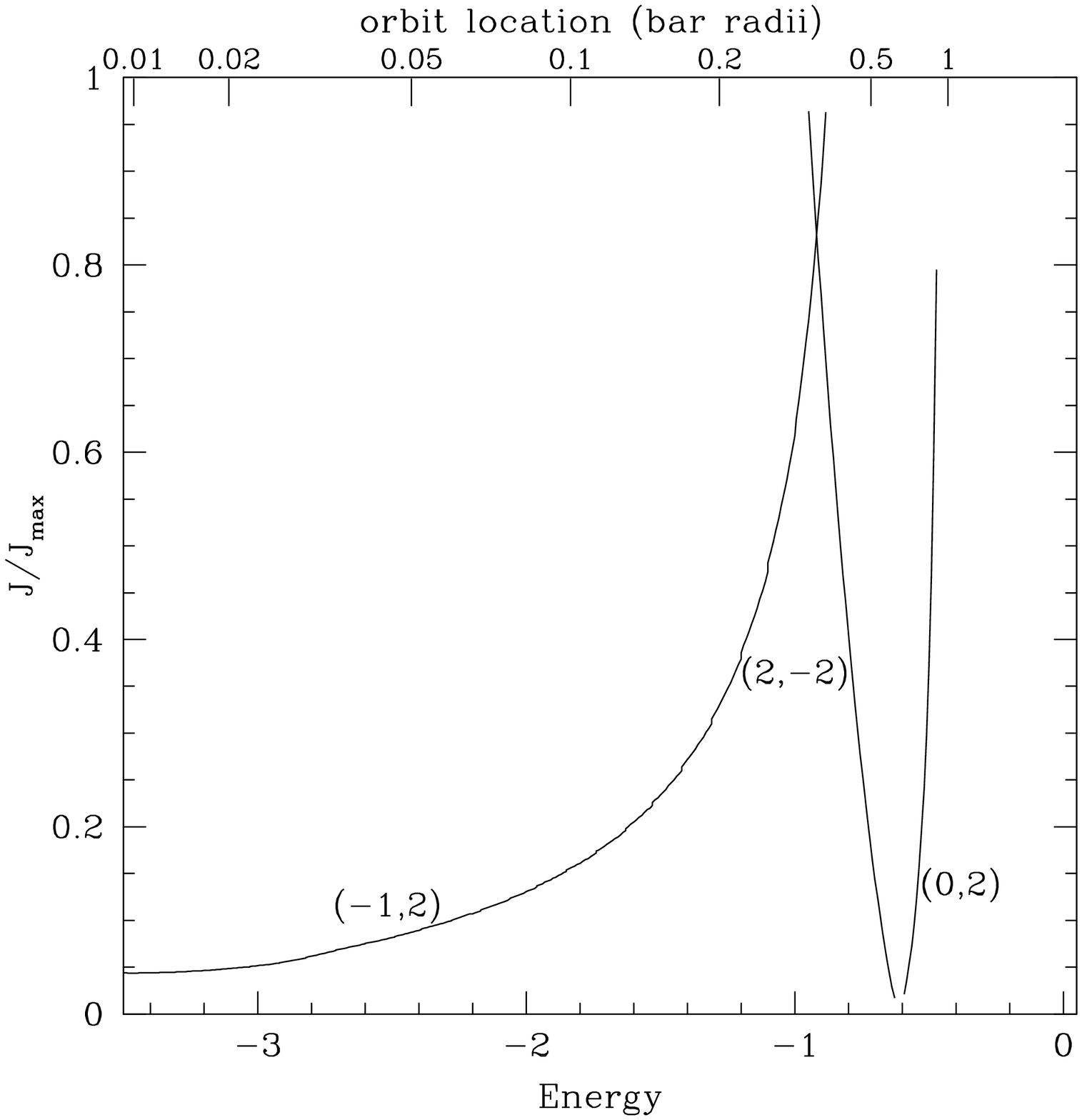}
  \caption{Left: the halo response and bar position during the evolution
    of an NFW-model halo.  The mean density is subtracted and the
    amplitude of the resulting wake coded from underdensity (white) to
    overdensity (black). Right: location of low-order resonances in
    energy (lower axis) and characteristic radius (upper axis) for the
    bar in the NFW profile.  The vertical axis describes the orbital
    angular momentum $J$ in units of the maximum angular momentum for
    a given energy, $J_{max}$.  The inner-Lindblad-like resonance
    extends throughout the inner cusp.  This resonance is absent from
    many dark-halo models with cores.}
\label{fig:haloreswake}
\end{figure}

It is easy to estimate that a strong bar can have important
consequences for halo evolution.  A toy model for a rotating
gravitational quadru\-pole is two masses in orbit at the same distance
from the center of a galaxy but at opposite position angles.  We can
use the dynamical friction formula to estimate the time scale for
transferring all of the bar's angular momentum to the halo. The answer
is: a few bar rotation periods \citep[see][]{Weinberg:85}.  The total
angular momentum in the bar approaches that of the dark matter halo
within the optical radius and therefore we expect that this torque can
significantly change the inner halo profile.

This physical argument is gross simplification, however, and wrong in
detail.  The Chandrasekhar dynamical friction formula is derived by
considering the momentum transfer in scattering of stars by a
traveling body.  However, a quasi-periodic halo orbit encounters the
rotating perturbation many times.  If the precession frequency leads
or trails the pattern, the average net torque applied to this orbit
will be zero.  Kicks received at the exact commensurabilities between
the orbital and pattern frequencies individual orbits break the
symmetry and result in a density response that trails the bar.  We
illustrate this by an n-body simulation\footnote{The NFW halo is
  realized by $10^7$ equal mass particles and evolved using the
  expansion method described in \cite{Weinberg:99}.  See
  \S\ref{sec:method} for additional detail.}  of an NFW halo forced by
a rotating bar (Fig.  \ref{fig:haloreswake}, left).  This figure shows
the halo density distortion in response to the bar.  The bar's size
and phase are shown schematically.  The bar pattern leads the response
pattern and therefore the bar torques the halo.

The location in the halo of the peak angular momentum transfer depends
on halo profile itself in two ways.  First, the torqued orbits will be
near commensurabilities between the orbital frequencies and the
pattern speed.  These commensurabilities or {\em resonances} take form
$l_r\Omega_r + l_\phi\Omega_\phi = m\Omega_{bar}$ where the three
values $\Omega$ are the radial, azimuthal and pattern frequencies,
respectively, $l_r$ and $l_\phi$ are integers and $m$ is the azimuthal
multipole index.  For the $l=m=2$ multipole, we will denote a
particular resonance by the tuple $(l_r, l_\phi)$.  Low-order
(high-order) resonances have small (large) values of $|l_r|$ or
$|l_\phi|$.  Second, a net torque requires a differential in
phase-space density on either side of a particular resonance.  If the
bar is inside the halo core, the there will be very few nearby
resonances and the torque is diminished.  The dominant resonant orbits
are at or beyond the core radius.  It is the location of resonances
that is key to a large bar torque rather than the dark matter density
of the inner halo.  Conversely, if the bar is in the dark-matter cusp,
orbits near and inside the bar radius cover a large range frequencies
with low-order resonances deep in the cusp.  The torqued cusp orbits
move to larger radii, decreasing the cusp density and decreasing the
overall depth of the potential well.  Both of effects cause the cusp
to expand overall.  Thus, the formation of a bar will naturally
eliminate an inner cusp.

Because the bar--cusp couple depends on near-resonant dynamics,
simulations with very high resolution will be required to resolve the
dynamics, as previously described (\S\ref{sec:noisepower}).  The
simulation shown in Figure \ref{fig:haloreswake} was the end-point of
a suite of simulations of increasing particle number from $10^4$
through $10^7$.  Convergence using the expansion method was obtained
for particle number $\ga4\times10^6$.  Again, direct summation
approaches \citep[e.g.][GRAPE]{GRAPE:00} and tree codes
\citep{Barnes.Hut:86} provide adaptive dynamic range but will most
likely require higher particle number to obtain the same convergence.
Note that even if the resonances are not resolved, these simulations
will still exhibit significant torque.  Our suite of simulations show
that the overall torque {\em increases} as the particle number {\em
  decreases}!  These same simulations give us good agreement with
Chandrasekhar's formula.  Such agreement is not a good indication of
that one is observing the correct dynamics.  Conversely, Chandrasekhar
formula works well in simulations because resonances obliterated by
artificial diffusion and therefore is well represented by scattering.
In short, it is difficult to see resonant effects in n-body
simulations because diffusion rate is high for moderate number of halo
n-body particles.

\subsection{Cusp removal in NFW profiles}
\label{sec:NFW}

\subsubsection{Methodology}
\label{sec:method}

The evolution of the halo may be estimated analytically using a
perturbation expansion of the collisionless Boltzmann equation and
solution of the resulting initial value problem.  The zeroth-order
solution specifies the equilibrium galaxy and the first-order solution
determines the forced response of the galaxy to some perturbation.
The second-order solution determines the first irreversible change in
the underlying distribution.  For time scales much larger than an
orbital time, the transients can be made arbitrarily small.  Explicit
comparisons with n-body simulations suggest that this approximation is
acceptable even for a small number of orbital time scales.
Mathematical details can be found in \citet{Weinberg:85,Weinberg:01a}.

The n-body simulations are performed using a parallel implementation
of the algorithm described in \citet{Weinberg:99} using the Message
Passing Interface (MPI). This algorithm defines a set of orthogonal
functions whose lowest-order member is the unperturbed profile itself.
Each additional member in the series probes successive finer length
scales.  Because all scales of interest here can be represented with
a small number of degrees of freedom, the particle noise is low.

The initial conditions are a Monte Carlo realization of the exact
isotropic phase-space distribution function for the NFW profile,
determined by Eddington inversion \citep[see][Chap.
4]{Binney.Tremaine:87}.  The bar figure is represented by a
homogeneous ellipsoid with axis ratios 1:0.5:0.05.  We derive the bar
force from the quadrupole part of the bar's gravitational potential.
This rotating disturbance is turned on adiabatically over four bar
rotation times to avoid sudden transients.  The use of the quadrupole
term only ensures that the dark-matter halo remains in approximate
equilibrium as the bar perturbation is applied.  We choose the
corotation radius to be the NFW scale length and the bar radius is
chosen to be 0.5 scale lengths.  The results described below are only
weakly sensitive to this choice.

\subsubsection{Results}
\label{sec:results}

\begin{figure*}
  \plotfiddle{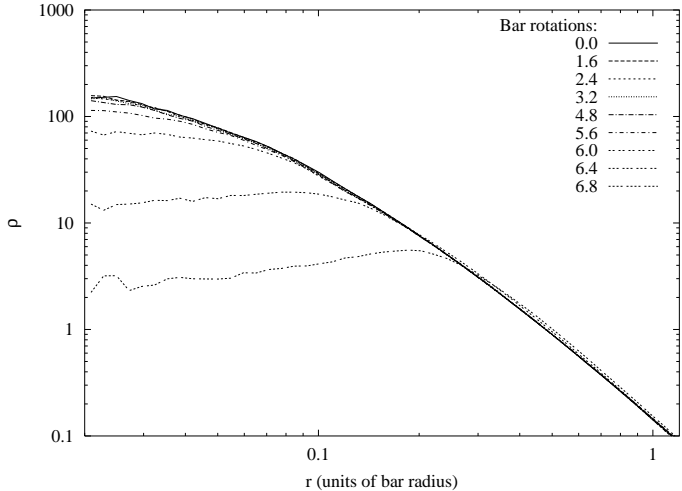}{225pt}{0}{75}{75}{-175pt}{-30pt}
  \caption{
    Evolution of an NFW profile with an embedded disk bar whose length
    is 1/2 of the scale parameter.  Profiles are a time sequence
    labeled by number of bar rotation times.}
  \label{fig:barNFW} 
\end{figure*}

The linear theory demonstrates that the angular momentum transfer to
the dark the matter takes place at commensurabilities between the bar
pattern speed and halo orbital frequencies and is dominated by a few
strong resonances.  The torque pushes the inner halo orbits higher
energy, removing gravitational support for the inner cusp. The halo
expands and flattens the cusp.  After a few bar rotation times
(several hundred million years) the central cusp is clearly flattened.
The n-body evolution, shown in Figure \ref{fig:barNFW}, is in
agreement with the analytic predictions up to about five bar rotation
times.  At this point, approximately 30\% of the available angular
momentum in the body rotation has been transferred to the halo.
Subsequent evolution is more rapid than predicted, presumably due to
the feedback of the evolving profile on the near-resonant orbits
although the details remain to be investigated; a similar super-linear
increase in torque was reported by \citep{Hernquist.Weinberg:92}.

The linear theory allows us to explicitly identify the dominant
resonances and the angular momentum transfer to halo orbits for models
both with and without cores.  The overall torque is dominated by one
or two resonances in each case.  For the NFW profile, the torque is
dominated by the resonance $(-1, 2)$; this is inner Lindblad resonance
analog.  This resonance does not occur for a comparable
\citet{King:66} model.  The first contributing radius for the King
model is $(2,-2)$ near the location of the core radius.  Higher-order
resonances occur at larger radii are well-confined to a single
characteristic radius (close to vertical in Fig.
\ref{fig:haloreswake}.  However, the $(-1,2)$ resonance is unique.  It
always exists in a cusp as $r\rightarrow0$ because
$\Omega_r\rightarrow 2\Omega_\phi$ as the orbital angular momentum $J$
approaches zero even though $\Omega_r$ $\Omega_\phi$ diverge for small
$r$.  Therefore there is always some value of $J$ for which $-\Omega_r
+ 2\Omega_\phi = 2\Omega_{bar}$.  For this reason, the $(-1, 2)$
resonance can affect orbits deep within a cusp, dramatically changing
the inner profile.  For a model with a core, the core expands somewhat
as angular momentum is transferred toward the outer halo but otherwise
remains qualitatively similar to its initial state.

\subsection{Discussion}
\label{sec:disc}

Bar instability is ubiquitous in self-gravitating disks.  Continued
accretion and cooling is likely to precipitate a bar instability in
the forming gas disk early on.  Cosmological simulation suggest that
these initial bars may be much larger than current-epoch bars.  Once
the rotating bar forms, its body angular momentum can be transferred
to the halo as described above, flattening and removing the inner
cusp.  It is possible for this process to occur in stages with
multiple gas bars; at each stage the inner core will grow.  There
will be a transition in the magnitude of the torque and halo evolution
when the inner $(-1,2)$ resonance is finally eliminated.

The underlying torque mechanism is not Chandrasekhar's dynamical
friction but resonant angular momentum transfer. However, the noise on
small scales will cause a high orbit diffusion rate with too few
particles.  The resulting scattering causes friction similar to that
predicted by using the dynamical friction formula.  Successful
simulation of the resonant torque requires in excess of $10^6$
particles in the dark matter halo and perhaps considerably more
depending on gravitational potential solver.  My tests show that the
torque is over-predicted in simulations with too few particles.

The presentation in \S\ref{sec:results} shows that the dominant
resonance driving the cusp flattening is low order.  Astronomical
sources of noise such as orbiting substructure, decaying spiral waves,
lopsidedness, etc.  produce large-scale deviations from equilibrium
that will not drive orbital diffusion in the inner halo within several
bar rotation times to affect these predictions\citep[see][for
estimates of these time scales]{Weinberg:01a, Weinberg:01b}.

As the disk matures and becomes stellar- rather than gas-dominated, a
normal stellar bar may form through secular growth or instability.
The first-generation bar evolution will have diminished the inner-halo
torque by flattening the cusp consistent with the
\citet{Debattista.Sellwood:00} arguments.  We conclude that the
observed lack of a central dark matter cusp in low surface brightness
galaxies and dwarfs is a consequence of simple dynamical evolution and
does not require a fundamental change to the nature of dark matter or
galaxy formation.  The difference in evolutionary end states may be
the result of strong star-formation feedback evacuating the more
weakly bound central potentials, lack of strong accretion events and
mergers after the primordial bar has disappeared, or a combination of
the two.  Indeed, such stochasticity naturally predicts the inferred
dispersion in present-day profiles.

\acknowledgements
This work was supported in part by NSF AST-9988146.

\bibliographystyle{astron}
% \bibliography{mnemonic,mybib,master,suppl}

\end{document}